\documentclass{emulateapj}

\shorttitle{$\it NuSTAR$ and Keck Observations of Heavily Obscured Quasars Selected by $\it WISE$}
\shortauthors{Wei Yan et al.}

\begin{document}

\newcommand{\Cahk}{\ifmmode \left[{\rm Ca H+K}\,\textsc{ii}\right\,\lambda3935,3968 \else Ca H+K\,$\lambda3935,3968$\fi}

\newcommand{\Mgb}{\ifmmode \left{\rm Mg}\,\textsc{i}\right\,\lambda5175 \else Mg\,{\sc i}\,$\lambda5175$\fi}

\newcommand{\hp}{\hphantom{,}}

\def\nar{\ref{New A Rev.}}          


\title{$\it NuSTAR$ and Keck Observations of Heavily Obscured Quasars Selected by $\it WISE$}


\author{Wei Yan\altaffilmark{1}}
\email{wei.yan.gr@dartmouth.edu}

\author{Ryan C. Hickox\altaffilmark{1}}

\author{Kevin N. Hainline\altaffilmark{2}}

\author{Daniel Stern\altaffilmark{3}}

\author{George Lansbury\altaffilmark{4}} 

\author{David M. Alexander\altaffilmark{5}}

\author{Raphael E. Hviding\altaffilmark{2}}

\author{Roberto J. Assef\altaffilmark{6}}
\author{David R. Ballantyne\altaffilmark{7}}

\author{Michael A. Dipompeo\altaffilmark{1}}
\author{Lauranne Lanz\altaffilmark{1}}
\author{Christopher M. Carroll\altaffilmark{1}}

\author{Michael Koss\altaffilmark{8}}

\author{Isabella Lamperti\altaffilmark{9}}

\author{Francesca Civano\altaffilmark{10}}
\author{Agnese Del Moro\altaffilmark{11}}

\author{Poshak Gandhi\altaffilmark{12}}
\author{Adam D. Myers\altaffilmark{13}}

\altaffiltext{1}{Department of Physics and Astronomy, Dartmouth College, Hanover, NH 03755, USA}

\altaffiltext{2}{Steward Observatory, University of Arizona, 933 North Cherry Avenue, Tucson, AZ 85721, USA}

\altaffiltext{3}{Jet Propulsion Laboratory, California Institute of Technology, 4800 Oak Grove Drive, Mail Stop 169-221, Pasadena, CA 91109, USA}

\altaffiltext{4}{Institute of Astronomy, University of Cambridge, Madingley Road, Cambridge CB3 0HA, UK}

\altaffiltext{5}{Department of Physics, University of Durham, South Road, Durham DH1 3LE, UK}

\altaffiltext{6}{N$\acute u$cleo de Astronom$\acute i$a de la Facultad de Ingenier$\acute i$a, Universidad Diego Portales, Av. Ej$\acute e$rcito Libertador 441, Santiago, Chile}

\altaffiltext{7}{Center for Relativistic Astrophysics, School of Physics, Georgia Institute of Technology, Atlanta, GA 30332, USA}

\altaffiltext{8}{Eureka Scientific, Inc., 2452 Delmer Street Suite 100,
Oakland, CA 946027 USA}

\altaffiltext{9}{Department of Physics and Astronomy, University College London, Gower Street, London WC1E 6BT, UK}

\altaffiltext{10}{Yale Center for Astronomy and Astrophysics, 260 Whitney Avenue, New Haven, CT 06520, USA}

\altaffiltext{11}{Max Planck Institute for extraterrestrial Physics, Giessenbachstrasse 1, 85748 Garching, Bayern, Deutschland}

\altaffiltext{12}{Department of Physics and Astronomy, University of Southampton, Highfield, Southampton SO17 3RT, UK}

\altaffiltext{13}{Department of Physics and Astronomy, University of Wyoming, Laramie, WY 82071, USA}



\begin{abstract}

A primary aim of the $Nuclear\; Spectroscopic \;Telescope\; Array$ ($\it NuSTAR$) mission is to find and characterize heavily obscured Active Galactic Nuclei (AGNs). Based on mid-infrared photometry from the Wide-Field Infrared Survey Explorer ($\it WISE$) and optical photometry from the Sloan Digital Sky Surveys, we have selected a large population of luminous obscured AGN (i.e., ``obscured quasars"). Here we report $\it NuSTAR$ observations of four $\it WISE$-selected heavily obscured quasars for which we have optical spectroscopy from the Southern African Large Telescope and W. M. Keck Observatory. Optical diagnostics confirm that all four targets are AGNs. With $\it NuSTAR$ hard X-ray observations, three of the four objects are undetected, while the fourth has a marginal detection. We confirm that these objects have observed hard X-ray (10--40 keV) luminosities at or below $\sim 10^{43}\; \rm erg \; s^{-1}$. We compare X-ray and IR luminosities to obtain estimates of the hydrogen column densities ($N_{\rm H}$) based on the suppression of the hard X-ray emission. We estimate $N_{\rm H}$ of these quasars to be at or larger than $10^{25}\; \rm cm^{-2}$, confirming that $\it WISE$ and optical selection can identify very heavily obscured quasars that may be missed in X-ray surveys, and do not contribute significantly to the cosmic X-ray background. From the optical Balmer decrements, we found that our three extreme obscured targets lie in highly reddened host environments. This galactic extinction is not adequate to explain the more obscured AGN, but it may imply a different scale of obscuration in the galaxy.

\end{abstract}

\keywords{galaxies: active -- galaxies: nuclei -- X-rays: galaxies}

\section{Introduction}

Quasars are particularly luminous examples of Active Galactic Nuclei (AGNs). Unobscured (type 1) quasars have been well-studied ever since they were discovered over 50 years ago (\citealt{haza63, schm63}). Thanks to their high luminosities, unobscured quasars dominate over host galaxy light at most wavelengths, making them relatively easy to observe and study. However, it is now known that half or more of quasars are obscured by gas and dust (e.g., \citealt{hick07abs, asse13wiseagn, mate17obsc}). The existence of many obscured (type 2) quasars has direct implications for the growth history of supermassive black holes (SMBH) in galactic centers (e.g., \citealt{alex12bh, hick18}). Type 2 quasars also have implications for the origin of the cosmic X-ray background (CXB; \citealt{gill07, trei09, ueda14cxb, aird15nustar}) and statistics of black hole growth across cosmic time. Some recent progress has suggested a large population of Compton-thick (CT) AGN with intrinsic column densities of $N_{\rm H}$ $ > 1.5 \times 10^{24}\; \rm cm^{-2}$ (e.g., \citealt{lans15nustarqso, ricc17merger, lanz18ct, marc18ctnustar}), as well as some contribution to the models of the CXB spectrum (e.g., \citealt{coma95, gill07cxb, trei09cxb, drap10, ueda14cxb}); however, this very obscuration makes the CT AGNs difficult to find and study. Therefore identifying these heavily obscured quasars becomes important for a general understanding of black hole evolution (e.g., \citealt{gand06xclust}).

\begin{figure*}[t]
\includegraphics[width=155mm]{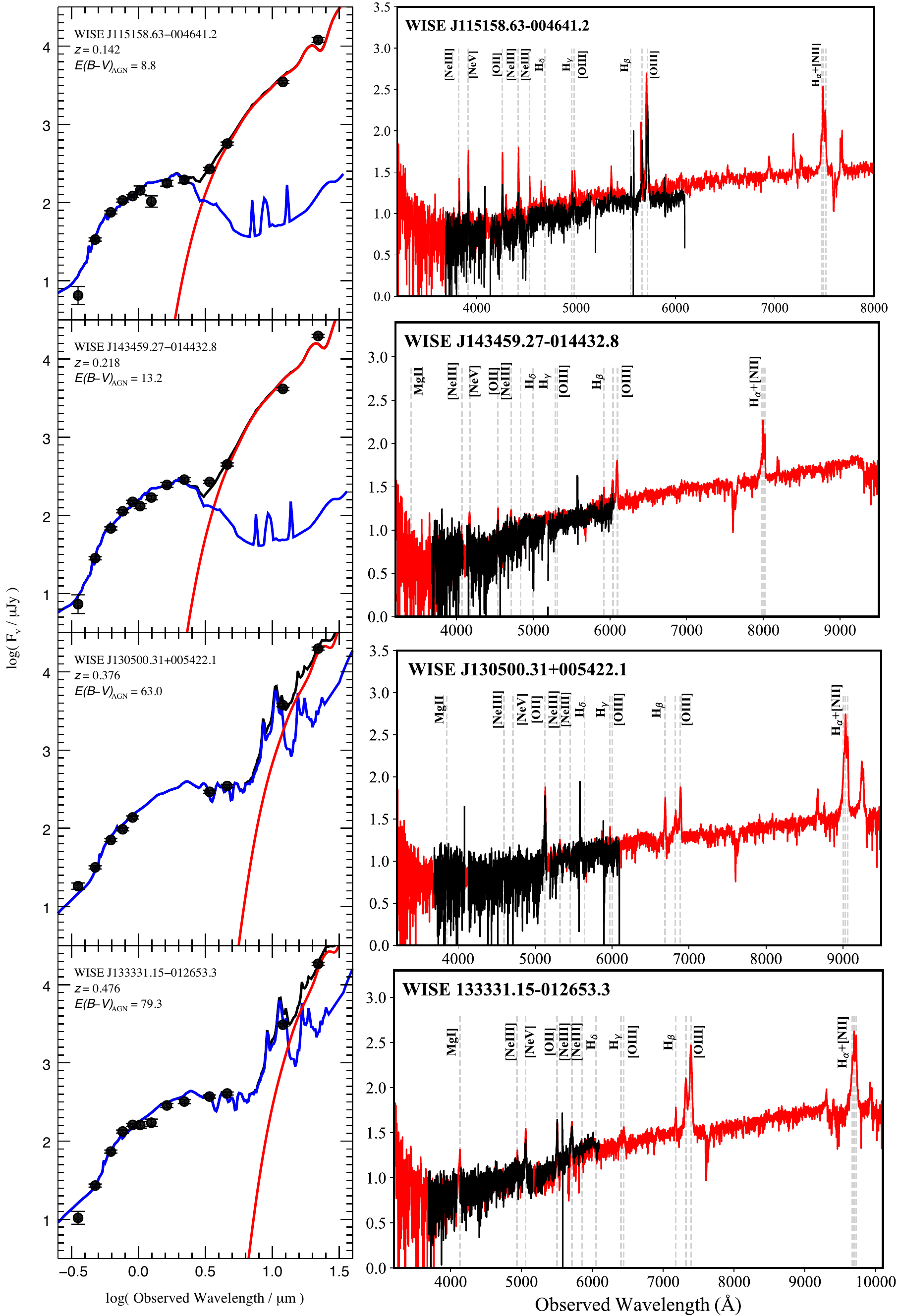}
\epsscale{1.1}
\caption{Spectral energy distributions (SEDs; $left$) and calibrated optical spectra from Keck (red lines) and SALT (black lines) in black solid lines respectively $(right)$. The SALT spectra are calibrated on the basis of Keck spectra by the level of the continuum and some bright emission lines (e.g., [\ion{O}{2}] and [\ion{Ne}{5}]). The SEDs and spectra indicate that all of our targets are AGN at moderate redshift, with moderate quasar-like luminosities ($L_{\rm bol} \sim 10^{45}$--$10^{46} \rm \; erg\; s^{-1}$) derived from their infrared spectra \citep{hain14salt}. In their SEDs, the blue line is galaxy template fitting and the red line is AGN fitting. The SEDs indicates significant nuclear dust extinction of $A_V\;>$ 20. Optical spectra from Keck show strong [\ion{O}{3}] emission and large Balmer decrements, indicating relatively large obscuration on galactic scale. \label{fig:optical}}
\end{figure*}

Mid-infrared (IR) observations, in particular with the $Spitzer\; Space\; Telescope$ and Wide-Field Infrared Survey Explorer ($\it WISE$), and X-ray surveys have now enabled us to detect significant samples of obscured quasars, as well as optical spectroscopy and photometry (e.g., \citealt{lacy04qso2, ster05, hick07abs, ster12wise, asse13wiseagn}). However, many properties of obscured quasars still remain unknown. For example, CT sources comprise a large fraction of lower-luminosity AGNs (e.g., \citealt{goul12comp, ricc17nustarwise}). While the CT fraction for more luminous quasars has important implications for their contribution to the CXB and galaxy evolution, few heavily obscured luminous quasars with intrinsic column densities over $10^{24}\; \rm cm^{-2}$ have been confirmed (e.g., \citealt{alex13nustar,gand14mrk34}).

X-ray observations of these heavily obscured quasars are challenging; for example, the photoelectric absorption cut-off (e.g., at around 10 keV for a $z = 0.2$ AGN absorbed by a column density of about $10^{24}\; \rm cm^{-2}$) dramatically reduces the flux of soft X-rays. This may bias the measured spectral parameters, such as the intrinsic power law photon index $\Gamma$ or $N_{\rm H}$, if fitting spectra with low or a limited energy range. Also, CT levels of absorption deeply suppress the primary continuum, revealing strong $\rm Fe\; K\alpha$ fluorescent line emission at 6.4 keV and a Compton reflection `hump' at $\sim$ 20-30 keV (e.g., \citealt{geor91}). 

Due to the limitations of observations, we have previously obtained only weak constraints on the distribution of the obscuring column density $N_{\rm H}$ of luminous quasars. For less powerful AGNs (i.e., Seyfert galaxies), the classic `unified model' is largely successful in explaining obscuration by varying viewing angles of the `torus' (\citealt{anto93, urry95, netz15unified}). However, it remains unclear whether this picture also holds for powerful quasars. The observed dependence of AGN obscuration on observed luminosity suggests a departure from the simplest unified model (e.g., \citealt{ueda03, simp05, trei10bhmerge, iwas12, buch15, asse15wiseqso}), and implicates and raises the possibility that a phase of heavy obscuration is important to process in galaxy evolution. The most powerful AGN may also be obscured by starbursts (e.g., \citealt{davi07agnsf, ball08agnsb}) or larger-scale gas clouds driven to the center of the galaxy by violent mergers or instabilities (e.g., \citealt{dima05qso, hopk08disk_aph, chen15qsosf, brus15agnoutflow, dipo18nlr}). Different quasar fueling mechanisms can produce different distributions of $N_{\rm H}$ and CT fractions (e.g., \citealt{drap12,  ball18cxb}). 

Due to recent deeper observations in the hard X-ray band ($>$ 10 keV) with the Nuclear Spectroscopic Telescope Array ($\it NuSTAR$; \citealt{harr13nustar}), we now can more accurately constrain the physical properties of obscured quasars (e.g., \citealt{lans17nustar}). As the first orbiting observatory to focus high-energy X-rays, $\it NuSTAR$ improves sensitivity by two orders of magnitude, as well as over an order of magnitude improvement in angular resolution relative to the previous generation of hard X-ray ($>$ 10 keV) observatories.

Based on mid-IR photometry from $\it WISE$ and optical photometry from the Sloan Digital Sky Survey (SDSS), we have selected 40 obscured quasars (e.g., \citealt{hain14salt, dipo15qsoz, dipo15wise}). Here, we use $\it NuSTAR$ to probe the X-ray obscuration in a sample of these quasars at $z < 0.5$. We report Keck and $\it NuSTAR$ observations of four $\it WISE$-selected heavily obscured quasars, for which \citet{hain14salt} presented optical spectroscopy from the Southern African Large Telescope (SALT). The paper is organized as follows: Section 2 details the sample selection; Section 3 describes the Keck data analysis; Section 4 describes the $\it NuSTAR$ data results.  The results are discussed and summarized in Section 5. Throughout the paper, we assume a $\Lambda$CDM cosmology with $H\textsubscript{o} = 69.6\; \rm km\; \rm s^{-1}Mpc^{-1}$, $\Omega \textsubscript{M} = 0.286$ and $\Omega\textsubscript{$\Lambda$} = 0.714$ \citep{wrig06coscal}.

\section{Obscured Quasar Sample}


\begin{figure}[t]
\includegraphics[width=90mm]{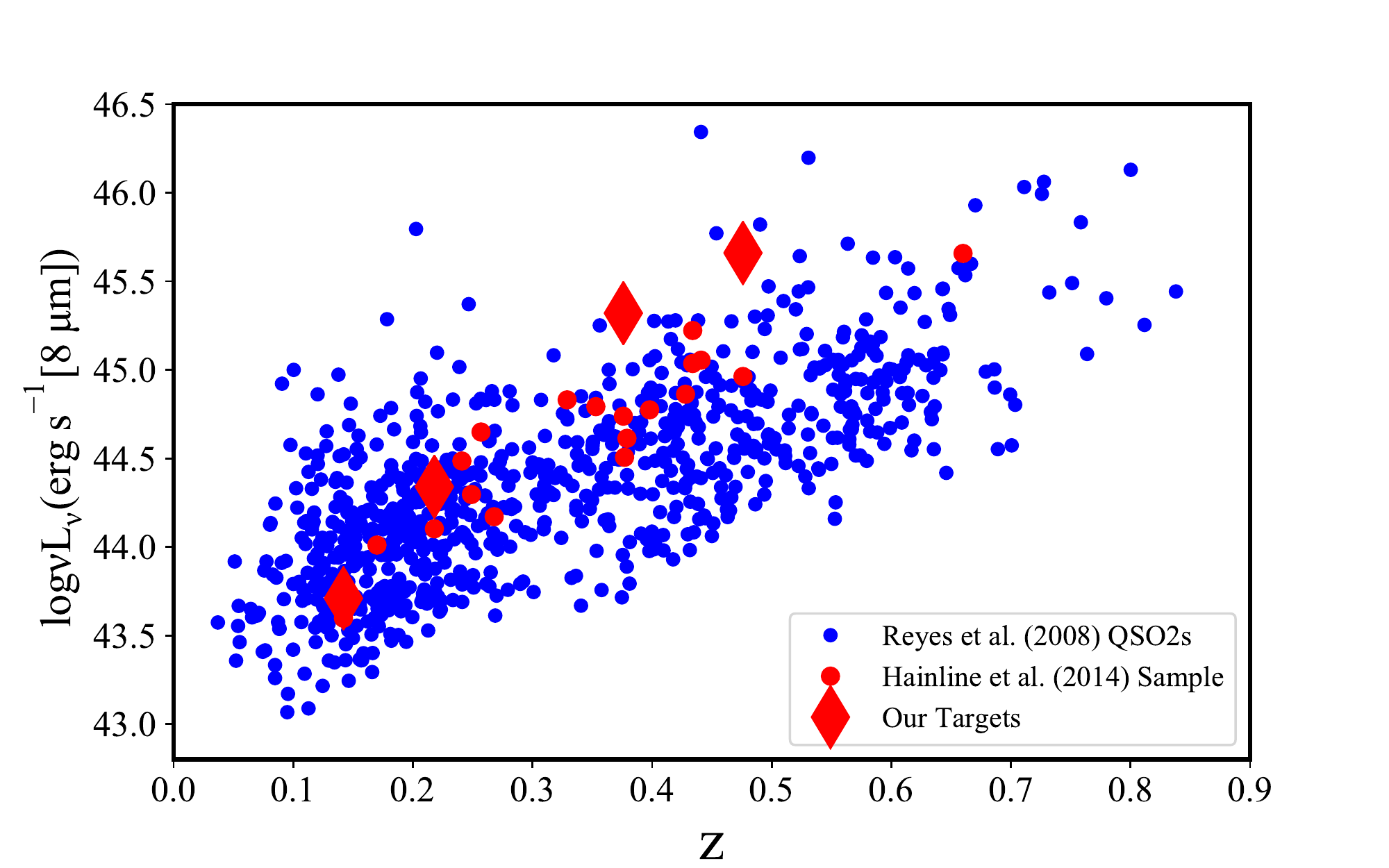}
\caption{ AGN luminosity at $8\; \mu m$ derived from SED fitting versus spectroscopic redshift for selected samples. We show the SDSS Type 2 quasar sample from \citet{reye08qso2} in blue dots. The parent sample (Group 1 in \citealt{hain14salt}) is shown as red dots and our four targets are shown as red dimonds. Two of our targets are AGNs with higher luminosities compared to those with similar redshifts. \label{fig:lum}}
\end{figure}

\begin{figure}[t!]
\includegraphics[height=95mm]{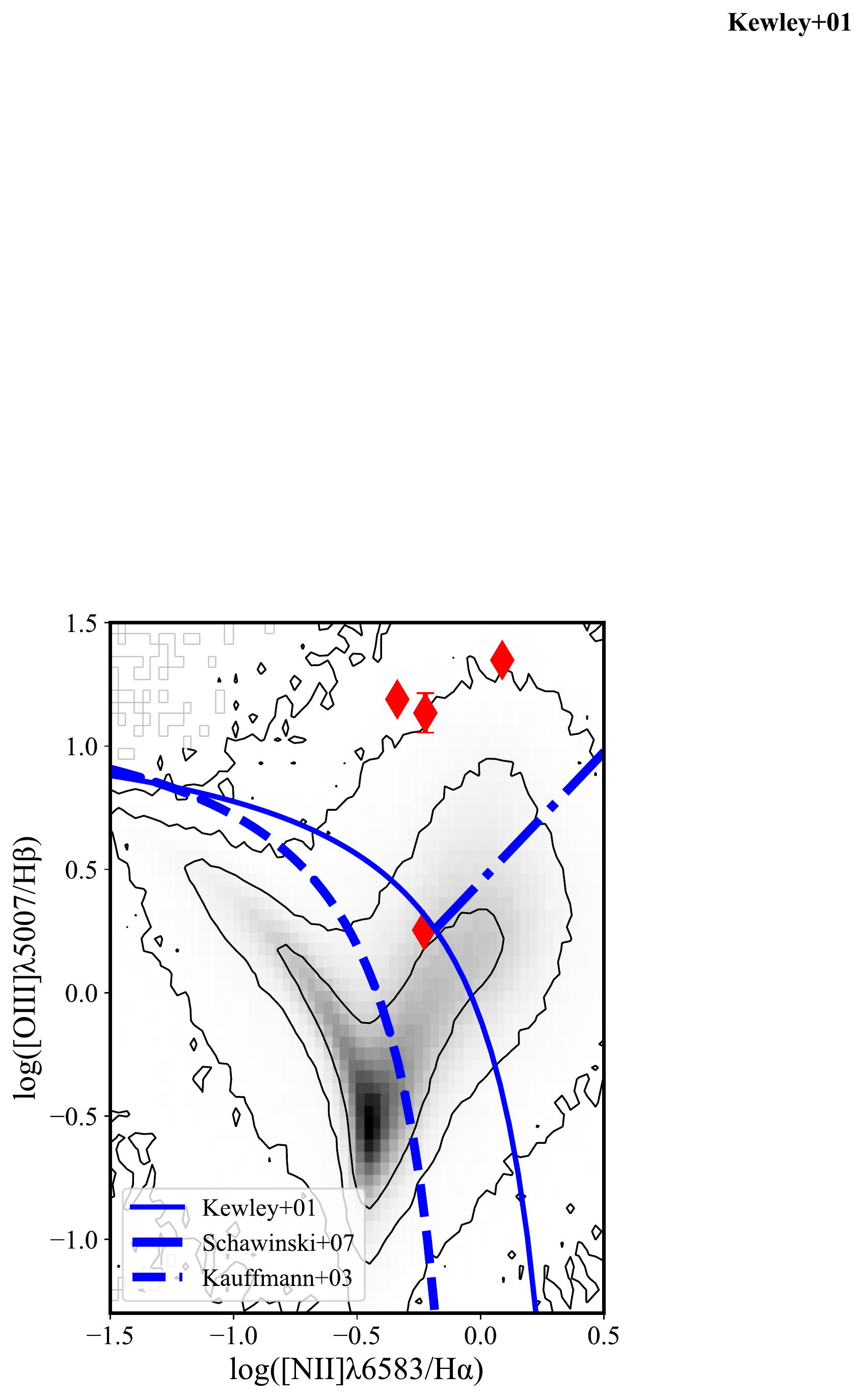}
\caption{Optical excitation diagnostic (BPT) diagram to separate AGN from star-forming galaxies using the fluxes from the Keck spectra. The solid blue curve is the theoretical boundary for the region occupied by starburst derived by the maximum starburst model from \citet{kewl01opt}. The dashed line is the empirical SF line from \citet{kauf03host}. The dot-dashed blue line is the empirical Seyfert–-LINER separation from \citet{scha07feed}. Red dots mark our targets. All of our targets lie in the AGN regime. For J1151-0046, $\rm H\beta$ line was not measured in the Keck spectra because of falling in a gap between the blue and red CCD. Therefore we use calibrated SALT flux for this line instead. \label{fig:BPT}}
\end{figure}

The parent population of 40 galaxies in total is drawn from \citet{hain14salt}, which were divided into two groups. Both groups are bright in the $\it WISE$ 22$\; \mu$m band (W4 $<$ 7; Vega)  and optical (20 $<\; g\;<$ 22) and use two slightly different $\it WISE$ selection criteria for identifying obscured quasars. Using both $\it WISE$ All-Sky and $\it AllWISE$ photometry, these were selected to have mid-IR and optical colors characteristic of obscured quasars (\citealt{hick07abs, ster12wise}). Objects of Group 1 are in the range $40 \arcdeg < RA <185 \arcdeg$, $-2\arcdeg < DEC < 2\arcdeg$ with $W1-W2 > 0.7$ and $7 \geqslant W4 \geqslant 6.5$, while those of Group 2 are $-2\arcdeg < DEC < 0\arcdeg$ with $W1-W2 > 0.8$, $W4\geqslant7.0$, and $r_{\rm AB}-W2_{\rm AB} > 3.1$. In the parent population, the density on the sky is $\sim\;$0.07$\;{\rm deg^{-2}}$. The luminosities of these $\it WISE$-selected objects lie in a range of $\log{(L_{8\;\mu{\rm  m}}/{\rm ergs^{-1}})} = 44.0-45.0$, with an average (median) of $\langle \log{(L_{8\;\mu{\rm  m}}/{\rm ergs^{-1}})} \rangle = 44.9 (44.8)$, while redshift has an average of $\langle z\rangle = 0.35$, with z $<$ 0.67.

The photometric criteria used in the parent population (\citealt{hain14salt, dipo15wise}) are relatively simple, and these sources are representative of large-scale, purely photometric statistical samples used, for example, for clustering studies to determine dark matter halo masses (\citealt{geac13qsocmb, dono14qsoclust, dipo14qsoclust, dipo15qsocmb, dipo17halomass}). The SALT optical spectroscopy has confirmed the presence of 40 AGN in the obscured quasar candidates observed in our program thus far (\citealt{hain14salt, hvid18qso2}), building on the success of an extensive spectroscopic survey of $\it Spitzer$-selected obscured quasar candidates \citep{lacy13spec}.  Our SALT quasars are at relatively low redshift ($z$ $<$ 0.5) and their infrared apectral energy distributions (SEDs) indicate moderate quasar luminosities ($L_{\rm bol} \sim 10^{45}$--$10^{46} \rm \; erg\; s^{-1}$).  These targets are generally considered as typical obscured quasars, however their gas obscuring columns and contributions to the CXB remain unexplored due to a lack of sensitive high-energy X-ray observations.

In this study, we have targeted four objects from Group 1 for which fits to the latest SEDs done in this work indicate significant nuclear dust extinction, $A_V\;>$ 20. These four targets J$133331.15-012653.3$, J$130500.31+005422.1$, J$143459.27-014432.8$ and J$115158.63-004641.2$ are broadly representative of the full \citet{hain14salt} sample, but with an emphasis on the most heavily dust-reddened objects to maximize the likelihood of identifying Compton-thick sources. Examining SDSS images of these four selected targets, we found that J$115158.63-004641.2$ is a late-stage merger, while the remaining three are isolated galaxies. In our latest SED fitting, for simplicity we modeled the extinction as a screen along the line of sight. To check that this is at least a broadly realistic representation of the AGN emission, we compared to other more sophisticated torus radiative transfer models (e.g., \citealt{silv04model, sieb15model}). We find that the broad shapes of our obscured AGN components are consistent with the output from these models for reasonable torus parameters, and with a similar dependence on optical depth. The SALT spectra in Figure~\ref{fig:optical} have been calibrated on the basis of Keck spectra by the level of the continuum and some bright emission lines (e.g., [\ion{O}{2}] and [\ion{Ne}{5}]). The optical spectra indicate that all the targets are AGN at $z<$ 0.5, with moderate quasar-like luminosities ($L_{\rm bol} \sim 10^{45}$--$10^{46} \rm \; erg\; s^{-1}$) according to the quasar luminosity function (e.g., \citealt{ueda03}), derived from their infrared SEDs \citep{hain14salt}. Since the SALT spectra of these targets do not cover Hydrogen lines, \citet{hain14salt} used the TBT (Trouille, Barger \& Tremonti; \citealt{trou11tbt}) criterion (see Figure 7 in \citealt{hain14salt}), which combines rest-frame g-z color with the emission line ratio [\ion{Ne}{3}]3869/[\ion{O}{3}]3726,3729. TBT disentangles AGN from star-forming galaxies and confirms that all our four targets are AGNs  \citep{trou11tbt}. Most of our targets clearly lie in the upper right area, which is the high-excitation part of the diagram and well into AGN regime.

Using the latest multi-component AGN and galaxy templates fit to the SEDs shown in Figure~\ref{fig:optical}, we determined the intrinsic unobscured luminosity of the AGN and the level of dust extinction. We have tried different template combinations and the systematic errors are small ($\sim$ 0.1 dex). These SED templates are the average results of the observations and are representative of the AGN population \citep{hick17sed}. The AGN templates include Assef type1 AGN template \citep{asse08sed} and \citet{rich06} comparison template. Our selection of galaxy templates includes Assef elliptical template, Assef spiral template, \citet{kirk15sfagn} star-forming galaxy template, and Assef irregular template. Compared to the SED fittings to the same targets in \citet{hain14salt}, here we adopt more types of the latest AGN and host galaxy templates and choose the best fits from these different combinations. This provides SED fitting that is less model-dependent and can be more easily compared to forthcoming work from our group (Carroll et al. in prep). From the best SED fits shown in Figure~\ref{fig:optical}, we obtained monochromatic intrinsic (unobscured and host corrected) infrared luminosities ($\nu L_\nu$) at rest frame $6 \rm \; \mu m$ and $8 \rm \; \mu m$ for comparison to different results in the literature, assuming that the mid-IR provides a reliable tracer of the intrinsic AGN luminosity.

Here we follow the strategies of \citet{ster14nustarwise} and \citet{lans14nustarqso, lans15nustarqso}, applied to our new $\it WISE$-selected obscured quasars. $\it NuSTAR$ has previously targeted specific samples of X-ray faint quasars selected through multiple techniques: luminous narrow-line ``Type 2 quasars" at $ z \sim 0.5$ selected with SDSS (\citealt{zaka03, reye08qso2, lans14nustarqso, lans15nustarqso}); extremely luminous obscured objects at $z \sim 2$ identified with very red colors in $\it WISE$ photometry (\citealt{eise12wise, ster14nustarwise, ricc17nustarwise}); and broad absorption-line (BAL) quasars identified through optical spectroscopy \citep{luo13balqso}. The characteristics of these samples vary from one to another, and it is unclear how these targets fit into the full underlying obscured quasar population. Therefore it is important to select more `typical' quasars, like our four photometrically-selected targets using infrared and optical data from $\it WISE$ and SDSS (e.g., \citealt{dipo15wise, hain14salt}).

We consider how our sample compares to the SDSS type-2 quasar sample, which is the basis for the previous estimate of the $N_{\rm H}$ distribution of obscured quasars \citep{lans15nustarqso}. We use the SDSS sample (with W1, W2, W3, signal-to-noise ratio (SNR) $>$ 3.0)  presented in \citealt{reye08qso2} (see also \citealt{zaka03}). The [\ion{O}{3}]$\lambda$5007 line is one of the indicators suitable for selecting a representative sample of obscured quasars to observe at X-ray energies. Since this emission line is one of the strongest visible in the optical at redshifts of about $z\; <$ 1, it potentially ensures that the AGN selection has unbiased intrinsic luminosity. \citet{reye08qso2} includes 887 objects covering 10,000$\; {\rm deg^{-2}}$. The density on the sky is about 0.08$\; {\rm deg^{-2}}$, similar to our parent WISE-selected sample. The sample has an average (median) of $\langle \log{(L_{8\;\mu{\rm  m}})}\rangle = 44.7 (44.3)$ while the upper limit of redshifts is set to be 0.84 ($\langle$z$\rangle = 0.324$, $z\textsubscript{median} = 0.279$). Hence the sample in \citet{reye08qso2} and our parent group are comparable. Figure~\ref{fig:lum} shows that our parent group is generally comparable with the larger sample in \citet{reye08qso2} in redshift and luminosity.

\section{Keck observations and results}

\begin{figure}[t]
\includegraphics[width=90mm]{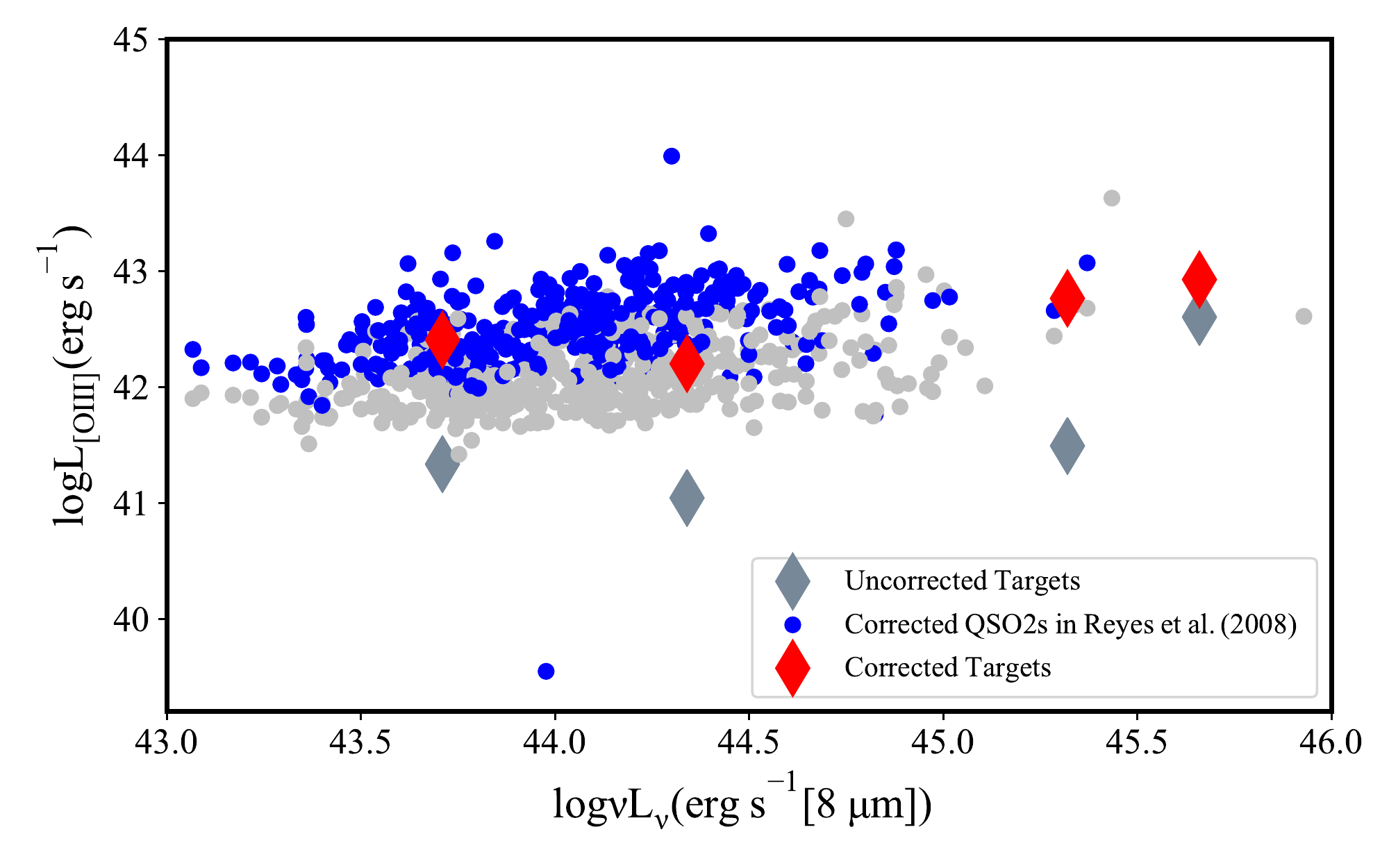}
\caption{[\ion{O}{3}] luminosity plotted against infrared luminosity for the objects in selected samples. We show the reddening-corrected [\ion{O}{3}] and infrared luminosities for the SDSS Type 2 quasar sample from \citet{reye08qso2} in blue solid dots, excluding those without measured Balmer emissions. Red diamonds are our targets with corrected [\ion{O}{3}] luminosities. Grey dots and diamonds are those before correction. After correcting for reddening, our targets and the \citep{reye08qso2} QSO2s lie in similar range of $L_{\text{ [\ion{O}{3}]}}$ and $L_{[8\mu m]}$. \label{fig:OIII_IR}}
\end{figure}

Since our SALT observations cover a limited wavelength coverage ($\sim$ 3680--6100 \AA), we obtained follow-up optical spectra with broader wavelength and higher SNR. The spectra with limiting spectrograph resolution ($\approx \sigma$ =100 km/s) were obtained during the night of 2017 April 28 (UT) at the W. M. Keck telescope with the Low Resolution Imaging Spectrograph (LRIS; \citealt{oke95keck}).  The exposure time varies from 300s to 450s.

\begin{figure}
\includegraphics[width=90mm]{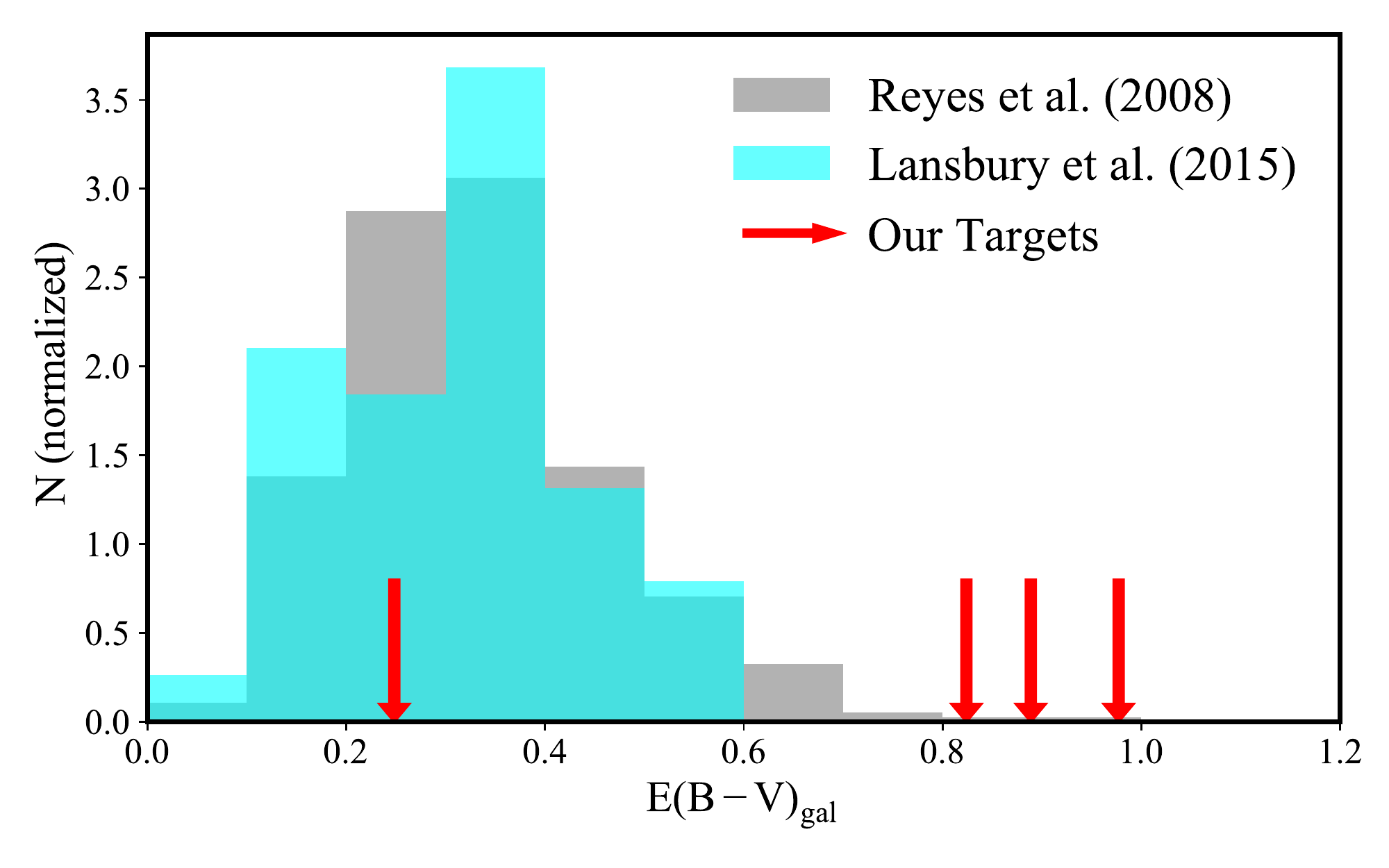}
\caption{Distribution of reddening ($E(B-V)$) of QSO2s in the narrow line region (NLR) at z $<$ 0.5, obtained from the Balmer decrements. The histograms are normalized. Grey columns are host enviroments of SDSS sample presented in \citet{reye08qso2}, while cyan ones are sources in \citet{lans14nustarqso}. Most of our targets lie around 0.8 $<E(B-V)<$ 1. In contrast, the extinction to the NLR of most QSO2s in previous samples is around 0.3, while our targets clearly show larger $E(B-V)$. This indicates that our targets are heavily-obscured AGNs with dusty host galaxies. \label{fig:E_dis}}
\end{figure}

\begin{deluxetable*}{lcccc}
\tabletypesize{\scriptsize}
\tablecaption{Optical emission lines observed with SALT and Keck.}
\tablewidth{0pt}
\tablehead{
\colhead{Object Name} & \colhead{$\rm {H\alpha}$} & \colhead{$\rm {H\beta}$} &\colhead{[OIII]} & \colhead{[NII]}
\\
\colhead{} & \colhead{$({\rm erg}\; {\rm s}^{-1}{\rm cm}^{-2})$} & \colhead{$({\rm erg}\; {\rm s}^{-1}{\rm cm}^{-2})$} & \colhead{$({\rm erg}\; {\rm s}^{-1}{\rm cm}^{-2})$} & \colhead{$({\rm erg}\; {\rm s}^{-1}{\rm cm}^{-2})$} 
}

\startdata

J$115158.63-004641.2$ & 1.96 & 0.26*& 4.07& 0.90 \\
J$143459.27-014432.8$ & 0.46 & 0.06 &  0.78 & 0.27 \\
J$130500.31+005422.1$ & 3.10 & 0.35 & 0.62 & 1.83 \\
J$133331.15-012653.3$ &0.79 & 0.21 & 4.61 & 0.97


\enddata 
\tablenotetext{*}{This flux is calculated based on calibrated SALT data rather than Keck.}
\label{table:emission}
\end{deluxetable*}

We adopt the PYSPECKIT software to model for the major emission lines at rest frame 3500--7000 \AA\ (e.g. [\ion{Ne}{5}], [\ion{O}{2}], [\ion{Ne}{3}], [\ion{He}{2}] 4686 \AA, ${\rm H}\beta$, [\ion{O}{3}], [\ion{O}{1}], ${\rm H}\alpha$, [\ion{N}{2}], [\ion{S}{2}]) following the general procedure in \citet{koss17specfit}.  We first fit steller emission using the penalized PiXel Fitting software ({\tt pPXF}; \citealt{Capp04}) and the templates from the Miles Indo-U.S. Catalog (MIUSCAT) library of stellar spectra \citep{vazd12miuscat}. The MIUSCAT library of stellar spectra contains $\approx$ 1200 well-calibrated stars covering the spectral region of 3525--9469\,\AA\ at a spectral resolution of 2.51\,\AA\ (full width at half maximum, hereafter as FWHM).  These spectra are used to compute with an initial mass function (IMF) slope of 1.3, considering a full range of metallicities (${\rm M}/{\rm H} = -2.27$ to $+0.40$) and ages (0.03--14 Gyr). As these templates are observed at higher spectral resolution (FWHM = 2.51\,\AA) than the AGN observations, they are convolved in {\tt pPXF} to the spectral resolution of each observation before fitting. When fitting the stellar templates all of the prominent, emission lines are masked.

Based on the Keck spectra, we use the BPT (\citealt{bald81bpt, kauf03host}) diagram, to confirm that our targets are all AGNs. The position of the BPT diagram is defined on the basis of [\ion{O}{3}]$\lambda$5007/${\rm H}\beta$, [\ion{N}{2}]$\lambda$6583/${\rm H}\alpha$, and [\ion{S}{2}]$\lambda\lambda$6716, 6731/${\rm H}\alpha$ flux ratios \citep{bald81bpt}, separating AGN from star-forming (SF) galaxies. In Figure~\ref{fig:BPT}, the solid blue curve is the theoretical boundary for the region occupied by starburst derived by the maximum starburst model from \citet{kewl01opt}. The dashed line is the empirical SF line from \citet{kauf03host}, the dot-dashed blue line is the empirical Seyfert--LINER (low ionization nuclear emission line regions) separation from \citet{scha07feed}. Compared to the AGN sample in \citet{reye08qso2}, we notice that our measured [\ion{O}{3}] luminosites are significantly lower (shown as gray in Figure~\ref{fig:OIII_IR}). This indicates that there may be heavy absorption and the narrow line emissions need further correction. For the Keck spectrum of WISE J115158.63-004641.2, the $\rm H\beta$ line fell in the gap between the blue and red CCD, and so was not measured. However, there are clear ${\rm H}\beta$ and [\ion{O}{3}] emission lines in the SALT spectrum (which we have otherwise not used for these line measurements because it covers a more limited wavelength range and does not have reliable absolute flux calibration). Therefore we use the SALT spectrum to obtain the ratio of the fluxes of the ${\rm H}\beta$ and [\ion{O}{3}] emission lines, by fitting Gaussian models to the two lines. We then multiply this ratio by the observed (and calibrated) [\ion{O}{3}] flux measured in Keck, to obtain an estimate of the flux in ${\rm H}\beta$ for this target. The positions of all four targets on the BPT diagram are shown in Figure~\ref{fig:BPT}. WISE J130500.31+005422.1 is on the AGN/composite boundary in the BPT diagram, but its ratio of [\ion{Ne}{3}]/[\ion{O}{2}] clearly identifies it as an AGN shown in \citet{hain14salt}. Therefore we further confirm that our targets are all AGNs. 

Since [\ion{O}{3}] can be used as the proxy of the intrinsic AGN power of obscured AGNs (e.g., \citealt{lama09oiii, vign10qso2}), we obtained [\ion{O}{3}] luminosities of our targets, shown in Figure~\ref{fig:OIII_IR}. We correct the narrow line ratios (${\rm H}\alpha$/${\rm H}\beta$), assuming an intrinsic ratio of R = 3.1 and the \citet{card89extinct} reddening curve. We use the model presented in \citet{char00balmer} to correct the absorbed [\ion{O}{3}] luminosities by Balmer decrements and derive the host galactic extinction. Since the error from the fit of the emissions are generally less than 0.1\%, the uncertainty of the corrected [\ion{O}{3}] flux mostly come from the correction of the reddening curve which is about 0.1 dex \citep{bosc02abs}. We then apply the same correction to our four targets and SDSS sample in \citet{reye08qso2}. Since both ${\rm H}\alpha$ and ${\rm H}\beta$ emissions are required for this correction, we exclude those in \citet{reye08qso2} without measured Balmer emissions in Figure~\ref{fig:OIII_IR}. After corrections, the [\ion{O}{3}] luminosities of our targets are comparable to those of the larger sample in \citet{reye08qso2}, whose mid-IR luminosities are derived from SED fitting as well. However, the correlation in Figure~\ref{fig:OIII_IR} needs to be used with caution since the parameters space may not be valid for hyper-luminous QSOs (e.g., \citealt{mart17wissh, hain16nlr}) or and the scattering is large at lower luminosities (e.g., \citealt{ueda15o3x}).

Figure~\ref{fig:E_dis} compares the distribution of the host galactic extinction between our targets determined from the Balmer decrements, the \citet{lans15nustarqso} sample and the \citet{reye08qso2} sample. Most of our targets have 0.8 $<E(B-V)<$ 1, while most of those in other two samples have 0 $<E(B-V)<$ 0.6. This indicates that, although we only selected our targets based on the extinctions in the nuclei, yet the host galaxies of those are heavily reddened. Therefore our targets are more obscured on larger scales compared to other narrow-line quasar samples (e.g. \citealt{reye08qso2, lans15nustarqso}), and these galaxies are among the dustiest galaxies found among typical X-ray selected AGN higher than almost all optically selected narrow line AGN (see Fig. 12 in \citealt{koss17specfit} for distribution). We will discuss the extinctions of the nuclei in the following section. 

We also attempted to obtain the velocity dispersion of our targets with Keck data. For the two most distant AGN, WISE J130500.31+005422.1 and WISE J133331.15-012653.3, the low S/N and relatively weak features prevent a velocity dispersion measurement. In WISE J115158.63-004641.2 and WISE J143459.27-014432.8, both measurements are consistent with being at or below the limiting spectrograph resolution ($\approx \sigma$ =100 km/s). Further higher spectral resolution measurements are needed for firmer constraints.


\section{$\it Nustar$ Observations and Results}

\begin{figure*}
\plotone{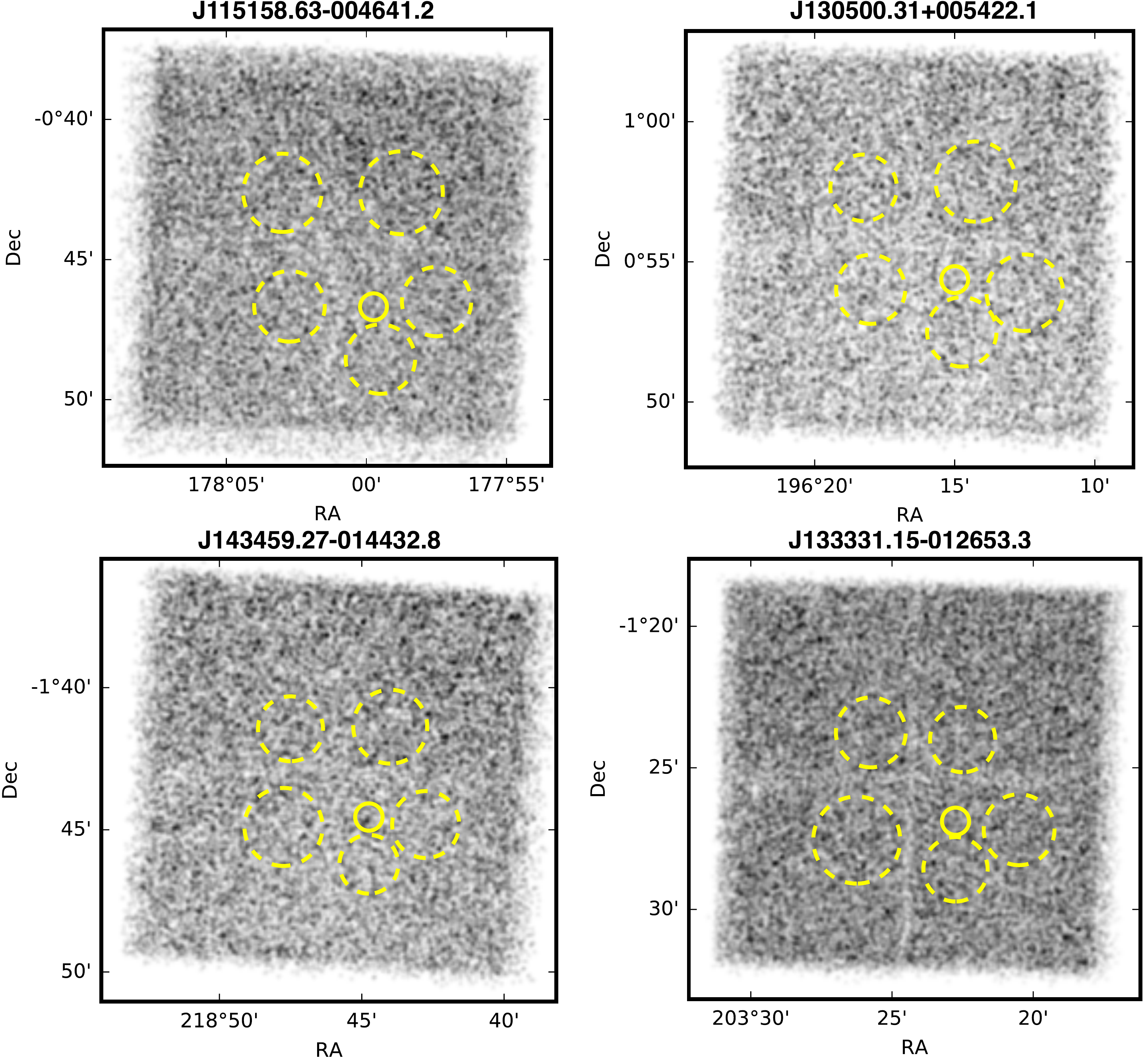}
\caption{$\it NuSTAR$ FPMA 3--24keV X-ray images of our four obscured quasar targets with source (solid) and background (dashed) regions shown. Only one target J$1434-0144$ (lower left) is detected. \label{fig:target}}
\end{figure*}

\begin{deluxetable*}{lccccc}
\tabletypesize{\scriptsize}
\tablecaption{$\it WISE$-Selected Obscured Quasar Targets.}
\tablewidth{0pt}
\tablehead{
\colhead{Object Name} & \colhead{$z$} & \colhead{$\log{L_{6\; \mu {\rm m}}}$} &\colhead{$\log{L_{\text{ [\ion{O}{3}]}}}$} & \colhead{$t_{\rm exp}$}&\colhead{$E(B-V)_{\rm AGN}$} \\
\colhead{} & \colhead{} & \colhead{$({\rm erg}\; {\rm s}^{-1})$} & \colhead{$({\rm erg}\; {\rm s}^{-1})$} & \colhead{$({\rm ks})$}&\colhead{} 
}

\startdata

J$115158.63-004641.2$ & 0.142 & 43.9$\pm$0.1 & 41.3& 40 & 8.8$\pm$1.5 \\
J$143459.27-014432.8$ & 0.218 & 44.5$\pm$0.3 & 41.0 & 36 & 13.2$\pm$1.8\\
J$130500.31+005422.1$ & 0.376 & 45.4$\pm$0.1 & 41.5& 26 & 63.0$\pm$9.7\\
J$133331.15-012653.3$ & 0.476 & 45.8$\pm$0.1 & 42.6& 24 & 79.3$\pm$12.0

\enddata 
\label{table:basic}
\end{deluxetable*}

\begin{figure}[t!]
\includegraphics[width=85mm]{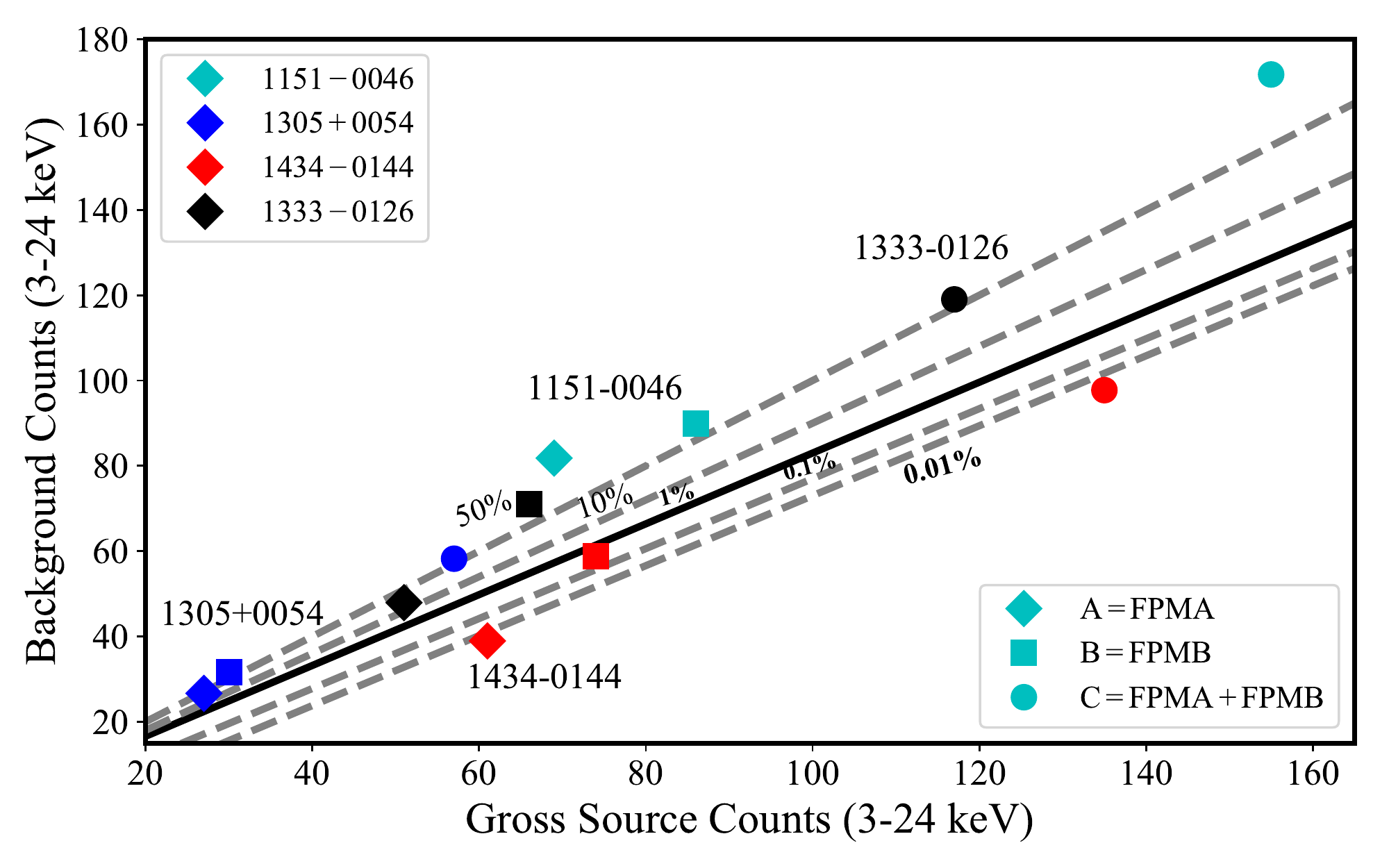}
\caption{$\it NuSTAR$ photometry (background counts versus gross source counts) at 3--24 keV for our four targets. Background counts are determined based on the count density in background regions scaled to the source aperture. We adopted different constant Poisson no-source probability (Weisskopf et al. 2007) as the solid and dashed lines. Only the $\it NuSTAR$ detected target J1434-0144 has a significant detection. 
\label{fig:count}}
\end{figure}

With $\it NuSTAR$ observations, we are able to probe obscured quasars in hard X-ray band, to obtain constraints of the absorption by gas. We obtained $\it NuSTAR$ images in each Focal Plane Module (FPMA and FPMB) of all four targets, the exposure times are listed in Table~\ref{table:basic}. To measure their brightness, we define regions of source and background as shown in Figure~\ref{fig:target}. We optimised the signal-to-noise ratio by using an aperture radius that encircles $\sim$ 50\% of the energy (29$\arcsec$; \citealt{harr13nustar}). We chose 4 to 5 background regions around each target that avoid the chip gaps. The radii of the these regions are chosen to be larger than that of the target, and some of them lie on different chips in the $\it NuSTAR$ FPMs from the target. We subtracted the backgrounds and calculate the net counts and errors of all 8 images in 3 bands (3--8 keV, 8--24 keV, 3--24 keV). The counting errors come from Poisson statistics. Net counts and errors are in Table~\ref{table:xray}. We note that instrumental properties vary for different chips on the $\it NuSTAR$ FPMs, so choosing background regions spanning different chips can, in principle, introduce some systematic uncertainty. However, we find that the surface brightnesses for the different background regions are consistent within Poisson errors, and choosing multiple large background regions allows us to minimize shot noise in estimating the background in the source region. 
We primarily focus on 3--24 keV because $\it NuSTAR$ reaches maximum effective area within this energy range and collects photons most efficiently, which is very important for faint targets like ours when there are few photons at higher energies. 


Following \citet{lans15nustarqso}, we calculated Poisson no-source probability (PB) at 3--24 keV for the four targets (Figure~\ref{fig:count}). We obtained a high PB for three of the four targets. Only the detected target J1434-0144 lies below $3\sigma$ probability after combining FPMA and FPMB.  For those, we take $3\sigma$ as the upper limits determined using the prescription of \citet{kraf03} following \citet{lans15nustarqso}. We convert the $\it NuSTAR$ count rate in the 3--24 keV band to luminosity in rest frame 10--40 keV band, assuming a power-law spectrum with $\Gamma = 1.8$. By assuming different levels of $N_{\rm H}$ distribution ($10^{22}$--$10^{25}\; \rm cm^{-2}$) and $\Gamma$ values (1.4--2.2), the conversion of the count rates into fluxes do not show significant differences once errors are taken into account. Therefore we assume a non-absorption case for the X-ray calculations. These are presented in Table~\ref{table:xray}.

To study the level of X-ray obscuration in these objects, we first consider the intrinsic relationship between infrared and X-ray defined for unobscured (type 1) AGNs in \citet{chen17}. We convert the 2--10 keV energy range used in \citet{chen17} into 10--40 keV luminosity again assuming a power-law spectrum with $\Gamma = 1.8$. Since we expect the targets to be heavily-obscured, the level of obscuration may reach the limit of models like $MyTorus$ \citep{murp09mytorus}, therefore we use $BORUS02$ model \citep{balo18borus} which is able to extend to $N_{\rm H}\;\sim\;10^{25.5} \; \rm cm^{-2}$ and determine the level of suppression of X-ray luminosities for column densities in the range between $1\times10^{24}\; \rm cm^{-2}$ and $3.1\times10^{25}\; \rm cm^{-2}$. In the overlapping column density range of validity ( $N_{\rm H}$ up to $10^{25.5} \; \rm cm^{-2}$), we obtain similar results from $BORUS02$ compared to $MYTorus$.

\begin{figure}[t]
\includegraphics[width=87mm]{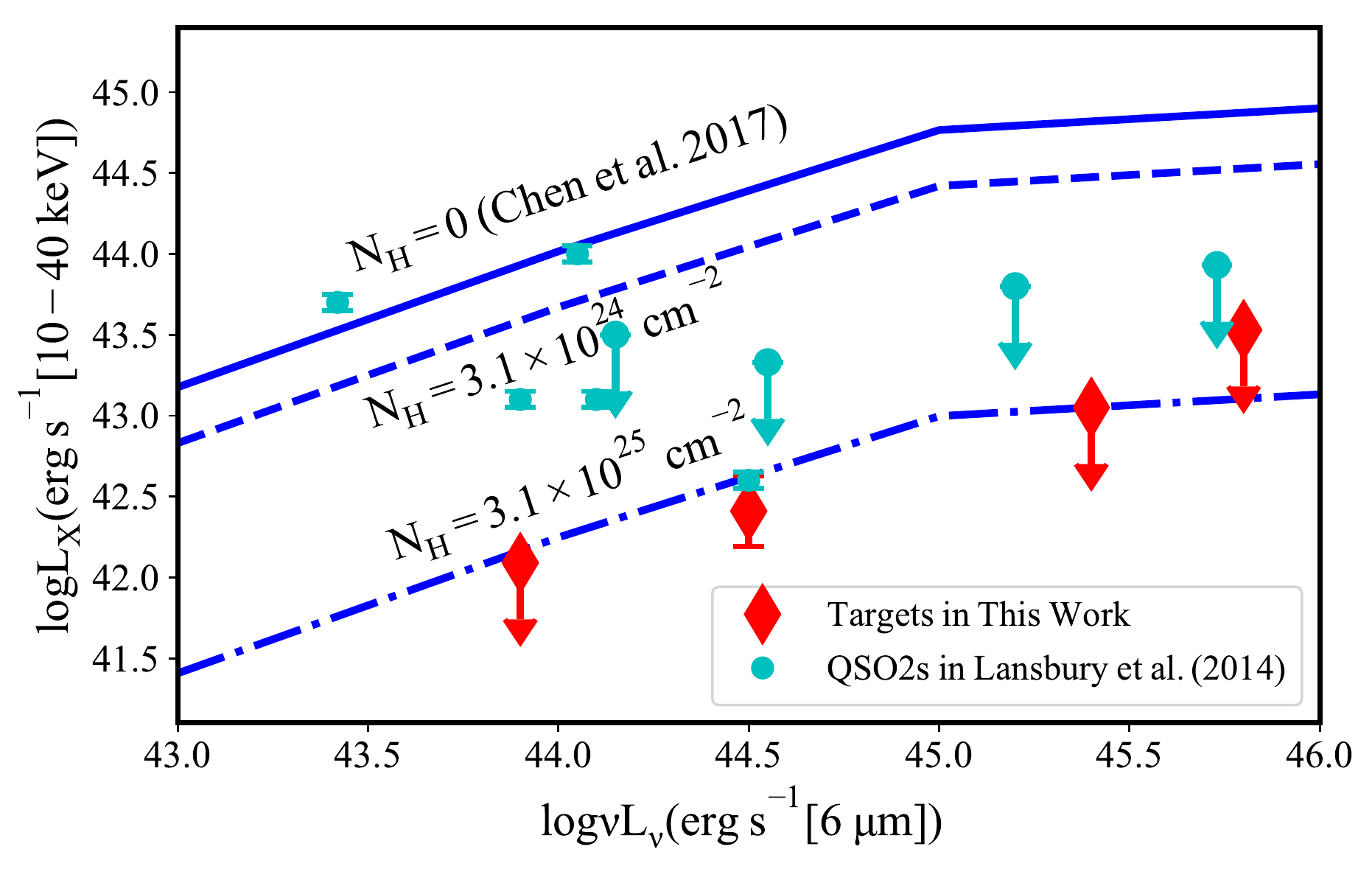}
\caption{X-ray luminosity vs infrared luminosity for obscured quasars. Red diamonds indicate our targets; Cyan dots indicate targets in \citet{lans15nustarqso}; the blue solid line is taken from a relationship  of \citet{chen17} derived from unabsorption AGN. The blue dashed line is derived with $BORUS02$ with a column density of $3.1\times10^{24}\; \rm cm^{-2}$, while the blue dot-dashed line is of $3.1\times10^{25}\; \rm cm^{-2}$. The low observed hard X-ray luminosities imply very heavy obscuration in all four targets. \label{fig:main}}
\end{figure}

\begin{figure}[t]
\includegraphics[width=87mm]{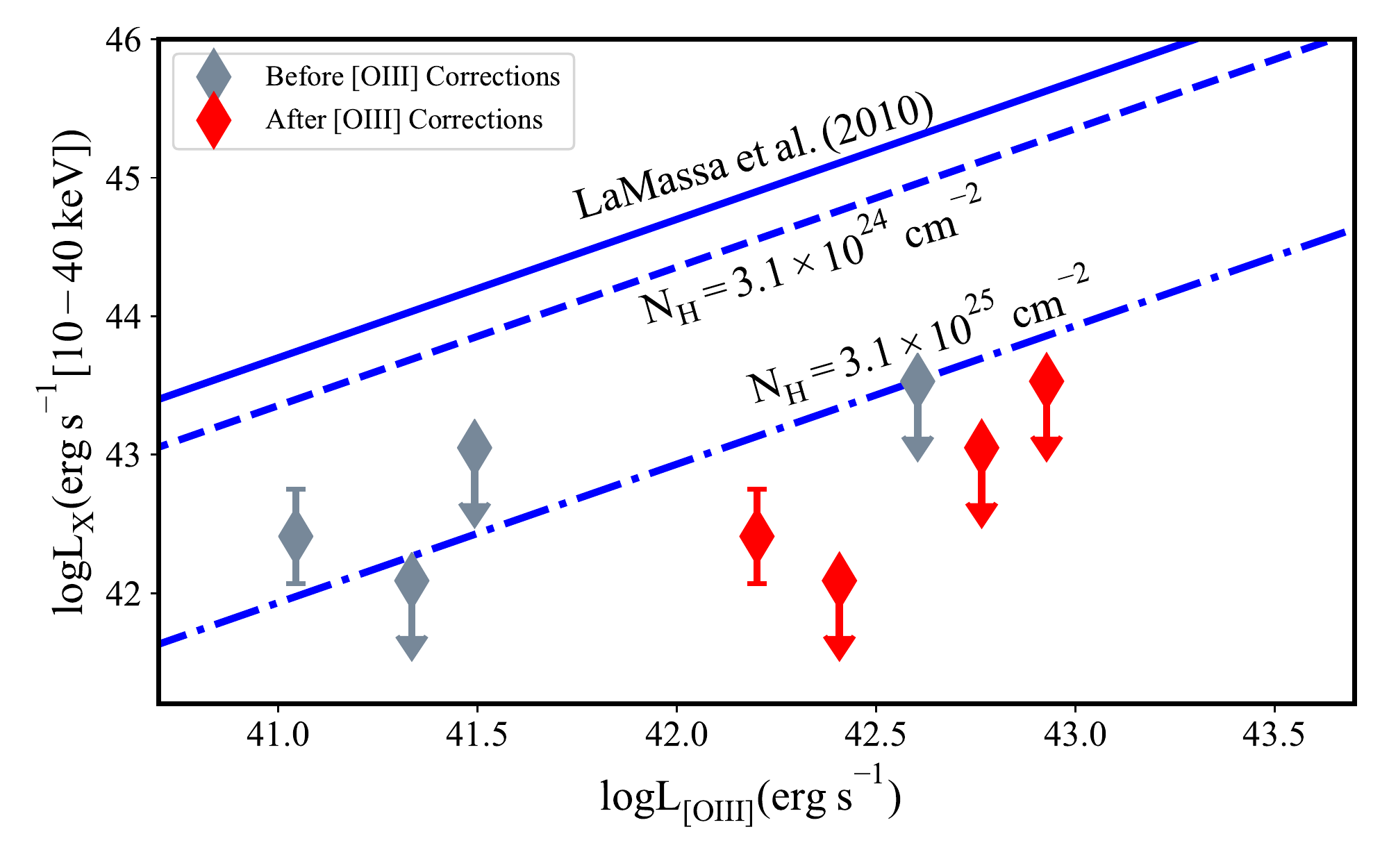}
\caption{X-ray luminosity vs [\ion{O}{3}] luminosity for obscured quasars. Grey diamonds indicate our targets before [\ion{O}{3}] luminosity corrections, while red diamonds indicate those with corrected [\ion{O}{3}] luminosities; the blue 
solid line is taken from a relationship derived from unobscured AGN in \citet{lama10OIII}. 
\label{fig:OX}}
\end{figure}

$BORUS02$ model is a new Monte Carlo code simulating radiative transfer (details will be presented in Balokovi{\'c} et al. in prep), which considers the interaction of the intrinsic X-ray continuum of AGN with the surrounding medium. The model is applicable to a wide variety of AGNs \citep{balo18borus}. Here we consistently take the photon index as $\Gamma = 1.8$ to obtain the relation between X-ray and infrared luminosities with different column densities \citep{lans15nustarqso}. Finally, we choose a redshift of $z = 0.3$ as the rough average redshift of our four targets. We set the inclination angle $\theta_{obs}$ between the line of sight and the symmetry axis of the torus as $ 60\; \arcdeg$ as the average value of the area integral of the torus. After obtaining the ratio of X-ray to infrared luminosities of all the targets, based on the suppression we estimate the range of column density where $N_{\rm H}$ of our targets fall. We then vary the column density along the line of sight in the range between $1\times10^{24}\; \rm cm^{-2}$ and $3.1\times10^{25}\; \rm cm^{-2}$ which is the upper limit for the models  , and derive the relationship between unabsorbed and absorbed luminosity for these column densities. Two representative relations with column densities of $3.1\times10^{24}\; \rm cm^{-2}$ and $3.1\times10^{25}\; \rm cm^{-2}$ are shown in Figure~\ref{fig:main}. Assuming a uniform density, to reach this level of obscuration, the mass would be unphysically high if distributed over a too large scale. Therefore obscuration is typically found on small scales for extreme column densities. \citep{hick18}

From Figure~\ref{fig:main}, it is clear that our objects have absorbed hard (10--40 keV) X-ray luminosities at or below $\sim 10^{43} \; \rm erg\; \rm s^{-1}$, with corresponding gas column densities $N_{\rm H}$ higher than $10^{25}\; \rm cm^{-2}$. We compute the ratios of X-ray to IR luminosities as a proxy for obscuration (e.g. \citealt{lans15nustarqso}). The statistical error of the infrared luminosity is small enough to be neglected. Based on the observed spread of mid-IR SED shapes for a population of obscured quasars (e.g., \citealt{hick17sed}), we estimate that the systematic error in our derived AGN luminosities due to the choice of SED template is about 0.1 dex. Since most of the X-ray luminosities are upper limits, it is possible that the actual column density value is higher than $10^{25}\; \rm cm^{-2}$. We repeat the same analysis above in different energy bands 3--8 keV and 8--24 keV and reach the same conclusions for the $N_{\rm H}$ estimate. By plotting the X-ray luminosities against the  [\ion{O}{3}] luminosities in Figure~\ref{fig:OX}, we show that our targets also have X-ray luminosities well below the relation taken from \citet{lama10OIII}. This confirms that our targets are heavily-obscured quasars with column densities above $10^{25}\; \rm cm^{-2}$. To be noticed, it is possible that at $z < 0.5$ about 17\% to 40\% optical-luminous quasars may have weak intrinsic X-ray luminosities \citep{luo13balqso}, and do not follow the relations shown in Figure~\ref{fig:OX}. However, since $N_{\rm H}$ values here agree with that derived from Figure~\ref{fig:main}, the low X-ray luminosities of our targets are more likely caused by heavy obscuration instead of weak intrinsic emission.

\begin{figure}
\includegraphics[width=90mm]{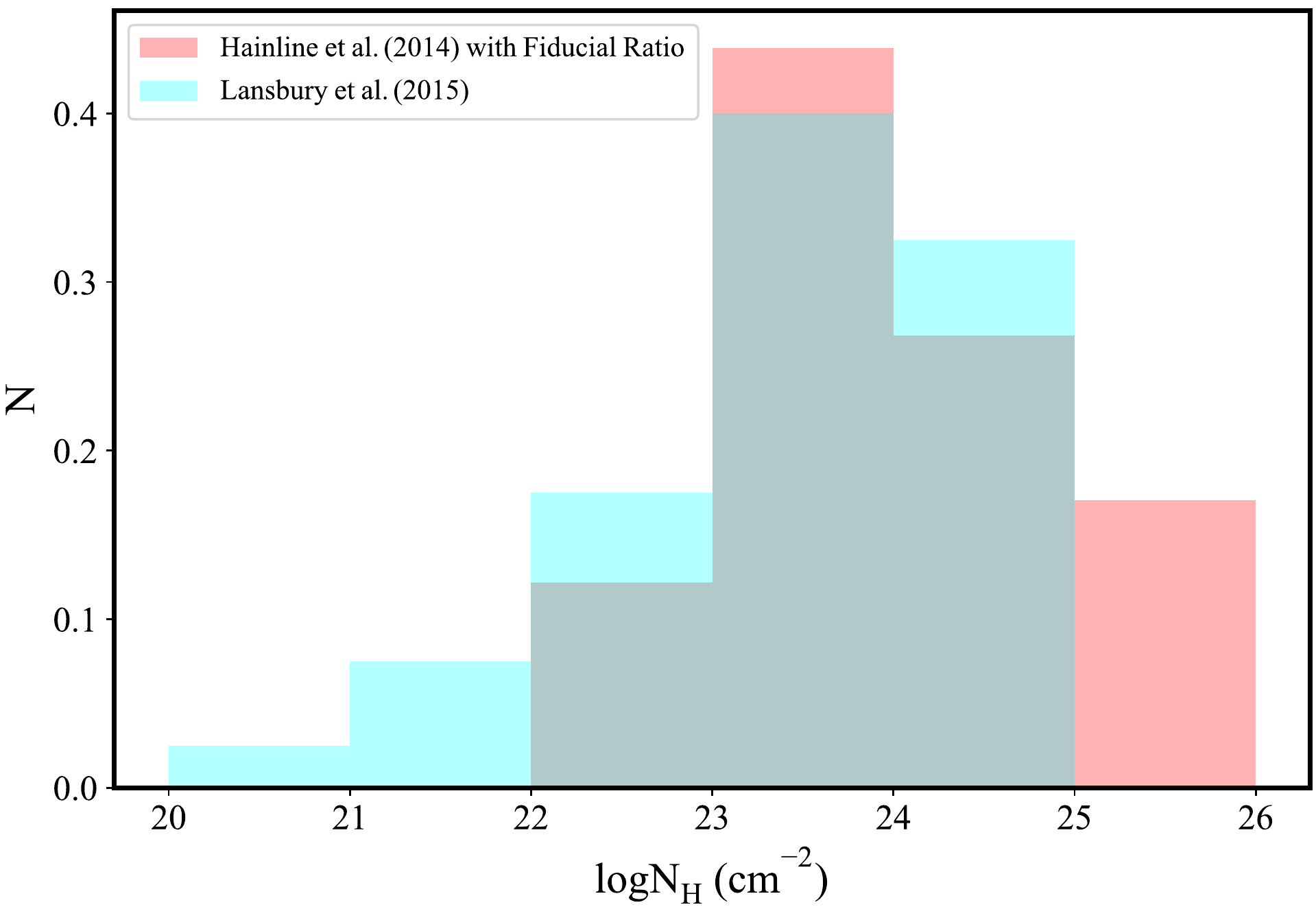}
\caption{The $\it NuSTAR$ $N_{\rm H}$ distribution of QSO2s at z $<$ 0.5. The red histogram shows the distribution for $\it WISE$-selected QSO2s in \citet{hain14salt}, assuming for the gas-to-dust ratio the conservative value of $N_{\rm H}$/$E(B-V)$ determined for our one $\it NuSTAR$-detected source. The cyan histogram shows the distribution for SDSS-selected QSO2s in \citet{lans15nustarqso}. With adopting the fiducial ratio, our $\it NuSTAR$ targets all lie in the range between 25 and 26, and the distribution is skewed toward very high $N_{\rm H}$. This indicates the potential existence of very heavily-obscured quasars with column densities above $10^{25}\; \rm cm^{-2}$ that may not have been identified in previous samples.
\label{fig:distribution}}
\end{figure}

Since the sample of SDSS in \citet{reye08qso2} and our parent group are comparable as discussed in Section 2, we can make a direct comparison between their $N_{\rm H}$ distributions. To estimate $N_{\rm H}$ of all QSO2s in our parent sample, we only have one $\it NuSTAR$ detected target, so we adopt its ratio of $N_{\rm H}$ to $E(B - V)$ as fiducial for the sample. As its column density lies close to $3.1\times10^{25}\; \rm cm^{-2}$, we take $3.1\times10^{25}\; \rm cm^{-2}$ for the estimation and $E(B - V)$ =13.2 which is derived from SED fitting. Therefore we adopt $2.3\times10^{24}$ as the value of `fiducial' gas-to-dust ratio. We then apply this value to 35 targets with known $E(B-V)$ values derived from our best SED fittings in this work and obtain an estimate of $N_{\rm H}$ of all these targets shown as red columns in Figure~\ref{fig:distribution}. Additionally we compare our $N_{\rm H}$ distribution of QSO2s from X-ray spectral analysis to that of SDSS-selected QSO2s in \citet{lans15nustarqso}.  From the comparison, our sample suggests a larger fraction of heavily obscured quasars at $N_{\rm H}\;>\;10^{25}\;\rm cm^{-2}$. As a result, the fraction of quasars that are Compton thick ($f_{CT}$) in our parent sample is (45.5$\pm$12.1)\% with fiducial ratio, while $f_{CT}$ in \citet{lans15nustarqso} is (32.5$\pm$11.5)\% which is $\sim$ 13\% (approximately 1$\sigma$) lower than ours. Considering the $f_{CT}$ predicted by nuclear starburst discs model (e.g., \citealt{gohi18fct}), our distribution with fiducial ratio tends to be closer to the large fraction of the model prediction ($f_{CT}$ around 60\% for quasars).

We also consider the gas-to-dust ratio of Small Magellanic Cloud (SMC) as well, since AGNs with cold absorption in X-ray spectra as well as at least two broad lines in optical/IR spectra tend to have a gas-to-dust ratio similar to the SMC \citep{maio01}. In the SMC, the ratio of gas to dust $N_{\rm H}$/$A_V$ is about $2.2\times10^{22}\; {\rm atoms\; cm^{-2}}\;{\rm mag^{-1}}$ (\citealt{bohl78hi, mart07fct, burt16absorption}). We parameterize  the ratio of $N_{\rm H}$/$E(B-V)$ based on typical reddening curves, i.e. SMC $R = A_V/E(B-V) =2.7$, and adopt $N_{\rm H}$/$E(B-V)$ of SMC as $6\times10^{22}\; {\rm cm^{-2}}$. We notice that SMC gas-to-dust ratio is about one magnitude lower compared to our `fiducial' ratio. Therefore if we adopt the SMC ratio instead of the `fiducial' ratio to estimate the $N_{\rm H}$ distribution of our parent sample, none of our selected sources have $N_{\rm H}$ above $10^{25}\; {\rm cm^{-2}}$. Since we have confirmed at least four targets have much higher obscuration level, we are underestimateing the Compton thick fraction with SMC dust-to-gas ratio. This indicates that SMC dust-to-gas ratio may not be applicable to quasars with the highest column density, i.e., one possibility is that dust sublimation on small scales in luminous AGNs causes relatively small dust column compared to the column of gas \citep{elit16dust}.

Our analysis suggests a large fraction of CT quasars. Although we use the single detected object to determine the column density in the ratio of  $N_{\rm H}$/$E(B-V)$, this detected X-ray source is the most conservative approach since other three targets might have a much larger column density. This points towards a possibility of an even larger fraction of heavily-obscured quasars and may shed light on the composition of the obscuring material in heavily obscured quasars. Meanwhile, it is also possible that we are indeed underestimating the dust reddening due to the fact that the high column density cannot be simply explained by the screen obscuration. In this case, a lower $E(B-V)$ which is obtained by our SED fitting methods would lead to a higher observed ratio of gas to dust.

\section{Discussion and Summary}

We report $\it NuSTAR$ hard X-ray observations of four targets belonging to a large population of 40 heavily obscured quasars selected by $\it WISE$ based on mid-IR photometry. All the targets are quasars with $L_{\rm bol}$ around $10^{45}\; {\rm erg\; s^{-1}}$ at $z<$ 0.5. The optical spectroscopy is from SALT and Keck, while the X-rays observations from $\it NuSTAR$. Three out of four objects are too faint to be detected in hard X-rays with $\it NuSTAR$, while the other one has only a marginal detection. From the upper limit and net counts of our X-ray observations in different bands, we confirm that our objects have observed X-ray luminosities at or below $\sim 10^{43}\; {\rm erg\; s^{-1}}$. This corresponds to gas column densities at or above $10^{25}\; {\rm cm^{-2}}$ derived from $L_{\rm X}$ and $L_{\rm 6\mu m}$, confirming that $\it WISE$ and optical selection can identify very heavily obscured quasars that may be missed in X-ray surveys. It also suggests the potential existence of many heavily-obscured AGN with column densities over $10^{25}\; \rm cm^{-2}$. The repeated Compton scattering in X-ray band makes it difficult to detect heavily-obscured AGNs (\citealt{wilm99abs, iked09xspec}). Although a small groups of CT sources have been found with X-ray surveys like XMM-COSMOS survey, yet only a few CT AGN candidates have been found with a large sample of sources (e.g., 67 out of 1855 in \citealt{lanz18cosmos}). Compared to Optical/mid-IR selection, X-ray selection requires specially designed spectral modeling \citep{lanz18cosmos} and are less efficient. Additionally, due to the limited photon statistics in deep fields, X-ray surveys are likely to miss AGNs with extreme $N_{\rm H}$ like our targets (e.g., \citealt{ueda14cxb, burt16absorption}). 

Considering torus model for AGN structure, the column densities imply more clumpiness of line-of-sight column densities (\citealt{dull05tori}; \citealt{nenk08torus}; \citealt{honi10clumpy}). Due to the heavy obscuration, it is difficult to constrain the parameters in the torus model. For example, we are only able to obtain the lower limits of the line-of sight column densities. This restraint may complex the model compositions which are not yet fully understood, including the covering factor and anisotropy of the torus (e.g., \citealt{netz15unified}). Additionally, while a large population of obscured AGNs could increase the total radiation density produced by SMBH in BH synthesis models, the heavily-obscured AGNs contribute little emission due to the extreme high $N_{\rm H}$ (e.g., \citealt{mart09agneff, nova13ct}). Given the large obscuration of our targets, the marginal detection and non-detections in the X-ray imply that these powerful AGNs do not contribute to the CXB for BH synthesis models.  \citep{hick18}.

Based on the narrow-line reddening of our targets, the extinctions of host galaxies are fairly high in three out of four. One of the target, WISE J130500.31+005422.1, is a late-stage merger with consistent large galactic absorption. Their Balmer decrements indicate substantial reddening in their host galaxies, which suggests the existence of different scales of obscuration in the heavily buried luminous AGNs. The galactic column densities are around $10^{23}\; {\rm cm^{-2}}$, compared to the nuclear $N_{\rm H}$ over $3.1\times10^{25}\; \rm cm^{-2}$ of these CT quasars. Although the heavily reddened host environment is still not adequate to explain the obscuration of these AGNs, yet this reasonably high galactic obscuration may suggest a connection between obscuration on different scales as may be associated with blackhole galaxy co-evolution (e.g., \citealt{hick18}).


\begin{deluxetable*}{lrrccccc}
\tabletypesize{\small}
\tablecaption{Net counts in 3--24 $\rm keV$ band and column densities of four targets.}
\tablewidth{0pt}
\tablehead{         
\colhead{Targets} 
& \colhead{Module A}  & \colhead{Module B} &
\colhead{Flux}
&\colhead{$logL_{10-40\; keV}$} 
&\colhead{Column Density}  & \colhead{$N_{\rm H}/E(B-V)$}
\\ 
\colhead{}  
& \colhead{Net Counts} & \colhead{Net Counts}
& \colhead{$(10^{-15}\; {\rm erg}\; {\rm s}^{-1}\; {\rm cm}^{-2})$} 
&\colhead{$({\rm erg}\; {\rm s}^{-1})$}
&\colhead{$(10^{25}\; {\rm cm}^{-2})$}  & \colhead{$(10^{24}\; {\rm cm}^{-2})$}
}
\startdata
  
  J$115158.63-004641.2$
	& $-7.2+7.3\tablenotemark{u}$  & $5.5+8.2^u$ 
	& $-1.1+7.8\tablenotemark{u}$
    & $ <42.1$ 
    & $>3.1$ & $>3.5$   
    \\

  J$143459.27-014432.8 $
	& $17.7\pm 7.9$ & $29.6\pm8.6$ 
	&$\hp18.3\pm9.2$ 
	& $42.4\pm0.2 $ 
	& 3.1 & $2.3\pm0.4$    
	\\

  J$130500.31+005422.1$ 
	& $-4.6+5.3\tablenotemark{u}$  & $-0.7+ 5.7^u$ 
	& $-3.1+4.2\tablenotemark{u}$ 
	& $<43.1$
	& $>3.1$ & $>0.5$   
	 \\
  
   J$133331.15-012653.3$
   & $-14.2+8.6\tablenotemark{u}$ & $5.9+9.4^u$ 
   & $-6.0+15.1^u$  
   &  $<43.5 $ 
   & $>1.0$ & $>0.1$





\enddata 

\tablenotetext{u}{Positive uncertainties denote upper limits.}
\label{table:xray}
\end{deluxetable*}

\acknowledgments

\begin{acknowledgments}

This research has made use of the NuSTAR Data Analysis Software (NuSTARDAS) jointly developed by the ASI Science Data Center (ASDC, Italy) and the California Institute of Technology (USA). The data presented herein were obtained at the W. M. Keck Observatory, which is operated as a scientific partnership among the California Institute of Technology, the University of California and the National Aeronautics and Space Administration. The Observatory was made possible by the generous financial support of the W. M. Keck Foundation. The authors wish to recognize and acknowledge the very significant cultural role and reverence that the summit of Maunakea has always had within the indigenous Hawaiian community.  We are most fortunate to have the opportunity to conduct observations from this mountain. Some of the observations reported in this paper were obtained with the Southern African Large Telescope (SALT). R.C.H.\ acknowledges support from NASA grant NNX15AP24G and National Science Foundation CAREER Award number 1554584. R.J.A. was supported by FONDECYT grant number 1151408.

\end{acknowledgments}

\bibliographystyle{apj}

 \newcommand{\noop}[1]{}

\end{document}